\documentclass[aps,twocolumn,superscriptaddress,showpacs]{revtex4-1}

\usepackage{graphicx}
\usepackage{dcolumn}
\usepackage{bm}
\usepackage{amsmath}
\usepackage{color}
\usepackage{ulem}
\usepackage{threeparttable}
\usepackage{enumerate}
\usepackage{paralist}

\begin{document}

\title{Anisotropic in-plane heat transport of Kitaev magnet Na$_2$Co$_2$TeO$_6$}

\author{Shuangkui Guang}
\affiliation{Department of Physics, University of Science and Technology of China, Hefei, Anhui 230026, People's Republic of China}

\author{Na Li}
\affiliation{Institute of Physical Science and Information Technology, Anhui University, Hefei, Anhui 230601, People's Republic of China}
\affiliation{Department of Physics, University of Science and Technology of China, Hefei, Anhui 230026, People's Republic of China}

\author{Qing Huang}
\affiliation{Department of Physics and Astronomy, University of Tennessee, Knoxville, Tennessee 37996-1200, USA}

\author{Ke Xia}
\affiliation{Department of Physics, University of Science and Technology of China, Hefei, Anhui 230026, People's Republic of China}

\author{Yiyan Wang}
\affiliation{Institute of Physical Science and Information Technology, Anhui University, Hefei, Anhui 230601, People's Republic of China}

\author{Hui Liang}
\affiliation{Institute of Physical Science and Information Technology, Anhui University, Hefei, Anhui 230601, People's Republic of China}

\author{Yan Sun}
\affiliation{Institute of Physical Science and Information Technology, Anhui University, Hefei, Anhui 230601, People's Republic of China}

\author{Qiuju Li}
\affiliation{School of Physics and Optoelectronics, Anhui University, Hefei, Anhui 230061, People's Republic of China}

\author{Xia Zhao}
\affiliation{School of Physics Sciences, University of Science and Technology of China, Hefei, Anhui 230026, People's Republic of China}

\author{Rui Leonard Luo}
\affiliation{Department of Physics and HKU-UCAS Joint Institute for Theoretical and Computational Physics at Hong Kong, The University of Hong Kong, Hong Kong, People's Republic of China}

\author{Gang Chen}
\email{gangchen@hku.hk}
\affiliation{Department of Physics and HKU-UCAS Joint Institute for Theoretical and Computational Physics at Hong Kong, The University of Hong Kong, Hong Kong, People's Republic of China}

\author{Haidong Zhou}
\email{hzhou10@utk.edu}
\affiliation{Department of Physics and Astronomy, University of Tennessee, Knoxville, Tennessee 37996-1200, USA}

\author{Xuefeng Sun}
\email{xfsun@ahu.edu.cn}
\affiliation{Institute of Physical Science and Information Technology, Anhui University, Hefei, Anhui 230601, People's Republic of China}
\affiliation{Department of Physics, University of Science and Technology of China, Hefei, Anhui 230026, People's Republic of China}
\affiliation{Collaborative Innovation Center of Advanced Microstructures, Nanjing University, Nanjing, Jiangsu 210093, People's Republic of China}

\date{\today}

\begin{abstract}

We report a study on low-temperature heat transport of Kitaev magnet Na$_2$Co$_2$TeO$_6$, with the heat current and magnetic fields along the honeycomb spin layer (the $ab$ plane). The zero-field thermal conductivity of $\kappa^a_{xx}$ and $\kappa^{a*}_{xx}$ display similar temperature dependence and small difference in their magnitudes; whereas, their magnetic field (parallel to the heat current) dependence are quite different and are related to the field-induced magnetic transitions. The $\kappa^a_{xx}(B)$ data for $B \parallel a$ at very low temperatures have an anomaly at 10.25--10.5 T, which reveals an unexplored magnetic transition. The planar thermal Hall conductivity $\kappa^a_{xy}$ and $\kappa^{a*}_{xy}$ show very weak signals at low fields and rather large values with sign change at high fields. This may point to a possible magnetic structure transition or the change of the magnon band topology that induces a radical change of magnon Berry curvature distribution before entering the spin polarized state. These results put clear constraints on the high-field phase and the theoretical models for Na$_2$Co$_2$TeO$_6$.

\end{abstract}

\maketitle

\section{Introduction}

Kitaev model is an exactly solvable spin model that supports Z$_2$ spin liquid with gapless Majorana excitations \cite{Kitaev}. In the presence of magnetic fields, the Kitaev model can further support gapped spin liquid with gapless and chiral Majorana mode that gives rise to half-quantized thermal Hall conductance. The presence of Kitaev interaction in the spin-orbit-coupled ${J=1/2}$ honeycomb magnets with $5d$ iridium ions (Ir$^{4+}$) \cite{Jackeli} and $4d$ ruthenium ions (Ru$^{4+}$) was suggested later \cite{Plumb}. Unfortunately, most of the relevant honeycomb irridates and even $\alpha$-RuCl$_3$ are well ordered at low temperatures \cite{NatureReviewPhysics}. $\alpha$-RuCl$_3$ in the magnetic field seems to support the half-quantized thermal Hall transport, that may be compatible with the gapped Kitaev spin liquid \cite{RuCl3-1, RuCl3-2, RuCl3-3}. Nevertheless, the actual ground state of $\alpha$-RuCl$_3$ in magnetic fields still needs further scrutiny. On the material's side, there were some efforts that proposed candidate Kitaev materials beyond the $4d/5d$ contexts, and these include the $4f$ rare-earth Kitaev magnets and the $3d$ transition metal Kitaev magnets. Along the line of the $3d$ transition metal Kitaev magnets, several honeycomb cobalt compounds such as Na$_2$Co$_2$TeO$_6$, Na$_3$Co$_2$SbO$_6$, and BaCo$2$(AsO$_4$)$_2$ were proposed and studied \cite{Liu1, Sano, Liu2, Zhong}.

The question about the Kitaev interactions in these Co-based materials, especially for BaCo$_2$(AsO$_4$)$_2$, has been raised in Refs. \onlinecite{Halloran, Das}. It was found that  the experimental results based on neutron scattering in BaCo$_2$(AsO$_4$)$_2$ can be consistently accounted for by the $XXZ$ model on the first and third neighbors but not by the more anisotropic spin model with Kitaev interactions and other pseudo-dipole interactions. It is then further remarked that, the frustration in BaCo$_2$(AsO$_4$)$_2$ is mainly from the competing first and third interaction, rather than from the Kitaev-related anisotropic interaction. Likewise, one could raise the same question for the honeycomb cobalt Na$_{2}$Co$_{2}$TeO$_{6}$ since no consensus has been reached concerning the microscopic models \cite{Songvilay, Lin, Kim, Samarakoon, Sanders, Yao1}. Here with our comprehensive thermal transport measurements, we are able to address this question, and the experimental results point to more anisotropic spin interactions.

In this work, we study the in-plane thermal conductivity and thermal Hall conductivity of Na$_{2}$Co$_{2}$TeO$_{6}$ with both the heat current and magnetic field along the $a$ or $a*$ axis. The magnetic field dependence of both thermal conductivity and thermal Hall conductivity display in-plane anisotropic behaviors. This is a strong indication of the intrinsic anisotropic nature of the spin interactions as a Kitaev material. In addition, the low-temperature $\kappa^a_{xx}(B)$ with $a$-axis field exhibits an anomaly at 10.25--10.5 T that is argued to be associated with an unexplored magnetic structure transition. We further provide a physical reasoning about the Berry curvature properties based on the contents of $\kappa_{xx}$ and $\kappa_{xy}$. This way of reasoning may be well generalized to other quantum magnets with the cooperative responses in $\kappa_{xx}$ and $\kappa_{xy}$ with respect to the external magnetic fields.

\section {Experiments}

High-quality single crystals of Na$_2$Co$_2$TeO$_6$ were grown by a flux method as previous reported \cite{Xiao}. Two thin-plate shaped crystals with size of 6.30 $\times$ 2.17 $\times$ 0.053 mm$^3$ and 5.20 $\times$ 2.10 $\times$ 0.047 mm$^3$ were used for heat transport measurements. The heat current and magnetic field were applied along the longest dimension of these two samples, which is the $a$-axis and $a*$-axis direction, respectively. The longitudinal thermal conductivity and thermal Hall conductivity were measured simultaneously by using the standard steady-state technique with ``one heater, three thermometers" \cite{Li}. Heat current and magnetic field were applied along the $a$ or $a*$ axis. The longitudinal and transverse temperature gradients were measured by three in-situ calibrated RuO$_2$ thermometers. The measurements were carried out in a $^3$He refrigerator equipped with a 14 T magnet.

\section{Results and Discussions}

\begin{figure}
\includegraphics[clip,width=6.5cm]{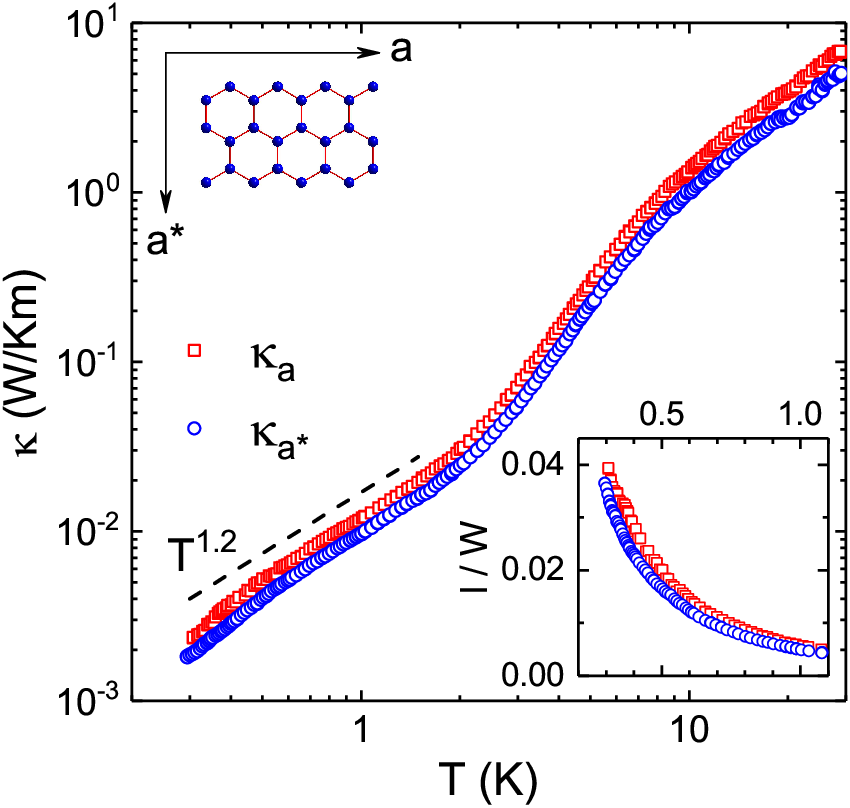}
\caption{Temperature dependence of the longitudinal thermal conductivity $\kappa^a_{xx}$ and $\kappa^{a*}_{xx}$ of Na$_2$Co$_2$TeO$_6$ single crystals, measured in zero magnetic field and with heat current along the $a$ axis (the zigzag direction) and the $a*$ axis (the Co-Co bond direction), respectively. The dashed line indicates a $T^{1.2}$ dependence of $\kappa$ at low temperature. The inset demonstrates the direction of $a$ axis and the $a*$ axis in the honeycomb layer of Co$^{2+}$, and the ratio of estimated phonon mean free path to the averaged sample width.}
\end{figure}

Figure 1 shows the temperature dependence of the longitudinal thermal conductivity $\kappa^a_{xx}$ and $\kappa^{a*}_{xx}$, measured in zero field and with heat current along the $a$ axis and the $a*$ axis, respectively. Na$_2$Co$_2$TeO$_6$ is known to have a zigzag \cite{Lefrancois, Bera} or triple-$q$ \cite{Chen, Lee} antiferromagnetic (AF) order below the N\'eel temperature ($T_N$ $\approx$ 27 K), followed by two possible spin re-orientation transitions around 16 K and 6 K. Though the $\kappa_{xx}(T)$ data shows no obvious anomaly around 27 K and 16 K, similar to the reported results \cite{Hong}, a slope change below 6 K can be related to the spin re-orientation. At subkelvin temperatures, the $\kappa_{xx}(T)$ exhibits a roughly $T^{1.2}$ behavior that is significantly weaker than the expected $T^3$ behavior of phonons and indicates the existence of magnetic scattering of phonons. Assuming the purely phononic contribution to thermal conductivity, the phonon mean free path can be estimated. As shown in the inset to Fig. 1, the ratios of the phonon mean free path to the averaged sample width are much smaller than 1 even at $T <$ 1 K, indicating the existence of microscopic phonon scattering effect at such low temperatures. It is notable that at zero field the $\kappa^a_{xx}$ and $\kappa^{a*}_{xx}$ have similar temperature dependence with slight difference in their magnitudes; whereas, they exhibit rather different magnetic field dependence for the in-plane fields.

\begin{figure*}
\includegraphics[clip,width=18cm]{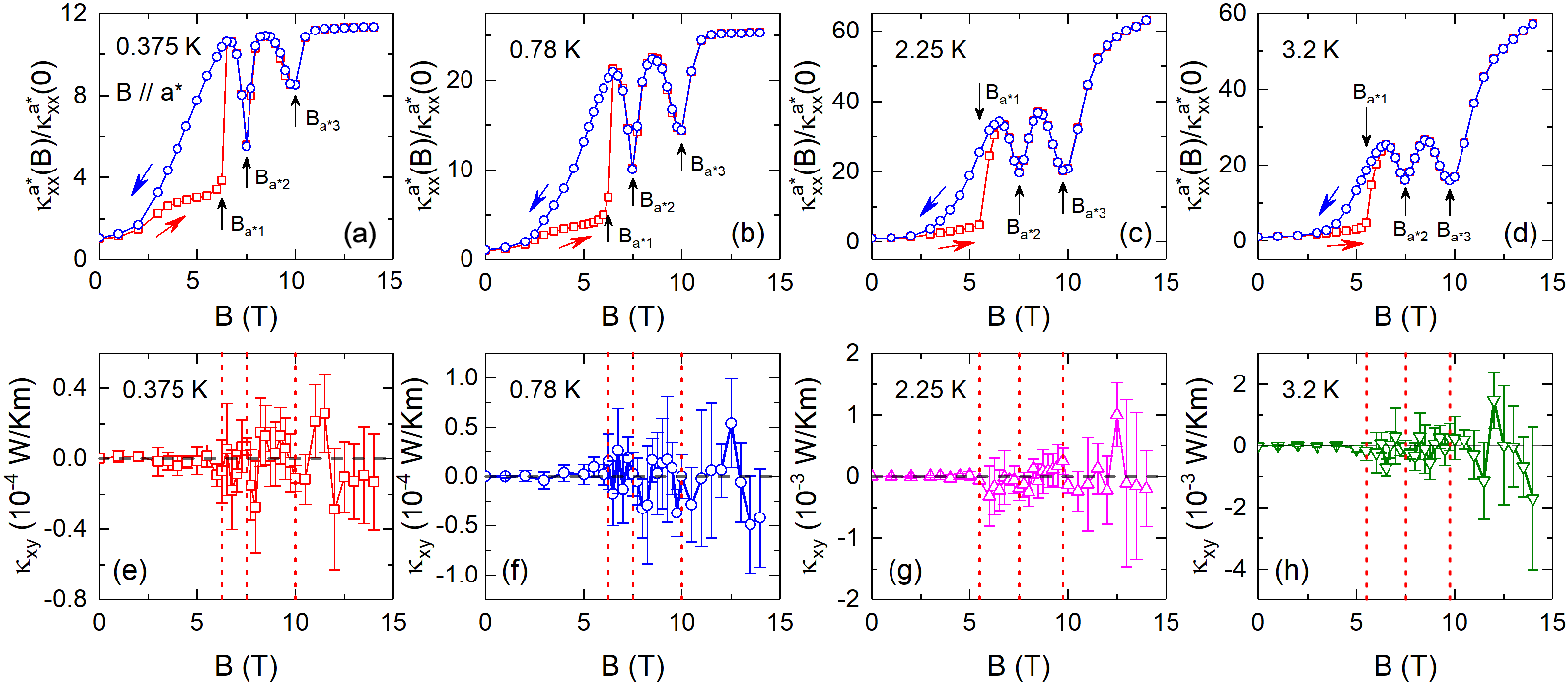}
\caption{Longitudinal thermal conductivity $\kappa^{a*}_{xx}$ and thermal Hall conductivity $\kappa^{a*}_{xy}$ of Na$_2$Co$_2$TeO$_6$ single crystals as a function of magnetic field with magnetic field and heat current applied along the $a*$ axis. (a-d) Isotherm curves of the $\kappa_{xx}$ with respect to the zero-field value as a function of magnetic field at 0.375 K, 0.78 K, 2.25 K and 3.2 K, respectively. The red and blue arrows indicate the process of applying magnetic field and the black arrows indicate the critical fields ($B_{a*1}$, $B_{a*2}$ and $B_{a*3}$) as mentioned in the main text. An obvious hysteresis can be observed at $B < B_{a*1}$. (e-h) Magnetic field dependence of thermal Hall conductivity $\kappa_{xy}$ at 0.375 K, 0.78 K, 2.25 K, and 3.2 K, respectively. Each data point is the averaged result from the repeated measurements for more than five times, the error bars represent the standard deviation. The red dot line indicates the positions of $B_{a*1}$, $B_{a*2}$ and $B_{a*3}$.}
\end{figure*}

Figures 2(a-d) show the magnetic field dependence of $\kappa^{a*}_{xx}$ with the heat current and magnetic field along the $a*$ axis at several selected temperatures from 0.375 to 3.2 K. It is found that $\kappa^{a*}_{xx}(B)$ shows a sharp jump at $B_{a*1} \sim$ 6.25 T with increasing field , and subsequently exhibits two minima at $B_{a*2} \sim$ 7.5 T and $B_{a*3} \sim$ 10 T before the saturation. According to the previous magnetization results \cite{Yao2}, the $B_{a*1}$ corresponds to a first-order magnetic transition and shifts to lower field with increasing temperature. Meanwhile, a large hysteresis occurs in $\kappa^{a*}_{xx}(B)$ at $B < B_{a*1}$. The rather sharp minima of $\kappa_{a*}(B)$ at $B_{a*2}$ and $B_{a*3}$ indicate two magnetic transitions, which is compatible with the recent inelastic neutron scattering measurements that revealed a field-induced intermediate magnetic state with partially polarized spins between 7.5 T and 10 T with magnetic field along the $a*$ axis \cite{Lin2}. All these results are consistent with our previous data measured at lower temperatures down to millikelvin temperatures \cite{Guang}.

Figures 2(e-h) show the planar thermal Hall conductivity $\kappa^{a*}_{xy}$ as a function of magnetic field at above selected temperatures. It should be pointed out that since the planar thermal Hall effect of Na$_2$Co$_2$TeO$_6$ was found to be rather weak, each data point displayed in these figures was the averaged value of data measured for more than five times. For all selected temperatures, there is no discernible $\kappa^{a*}_{xy}$ signal at $B < B_{a*1}$ and some sizeable thermal Hall signal appears at higher magnetic fields. At high temperatures of 2.25 K and 3.2 K, there is almost no signal between $B_{a*1}$ and $B_{a*3}$, but a sign change or weak drop appears at $B_{a*3}$ and there is a negative signal at $B > B_{a*3}$. Whereas, at lower temperatures of 0.375 K and 0.78 K, there is no discernable signal between $B_{a*1}$ and $B_{a*2}$, and a drop occurs at $B_{a*2}$ to reach a negative signal, followed by a sign change to positive signal before arriving at $B_{a*3}$; at $B > B_{a*3}$, there is the same behavior as that at high temperatures. Moreover, an obvious feature was observed in the high-field region above $B_{a*3}$. That is, the $\kappa^{a*}_{xy}$ signal first shows a peak with positive signal and then changes the sign at higher fields. This high-field feature is shared by the data at all selected temperatures, indicating an intrinsic narrow region with the positive $\kappa_{xy}$ before entering the high-field fully polarized state.

The planar thermal Hall conductivity and thermal conductivity was recently studied at higher temperature region of 2--30 K by Takeda {\it et al.} \cite{Takeda}. It is interesting to compare their results and ours. First, our $\kappa^{a*}_{xx}(B)$ data display much sharper minima at $B_{a*2}$ and $B_{a*3}$, indicating better sample quality. Second, although their low-field $\kappa^{a*}_{xy}(B)$ data are also almost zero, they show a sharp minimum with negative values at $B_{a*1}$. This feature is absent in our results. Third, their high-field $\kappa^{a*}_{xy}(B)$ data at 2 and 5 K also exhibits rather complicated behavior with sign changes, which is similar to ours.

\begin{figure*}
\includegraphics[clip,width=18cm]{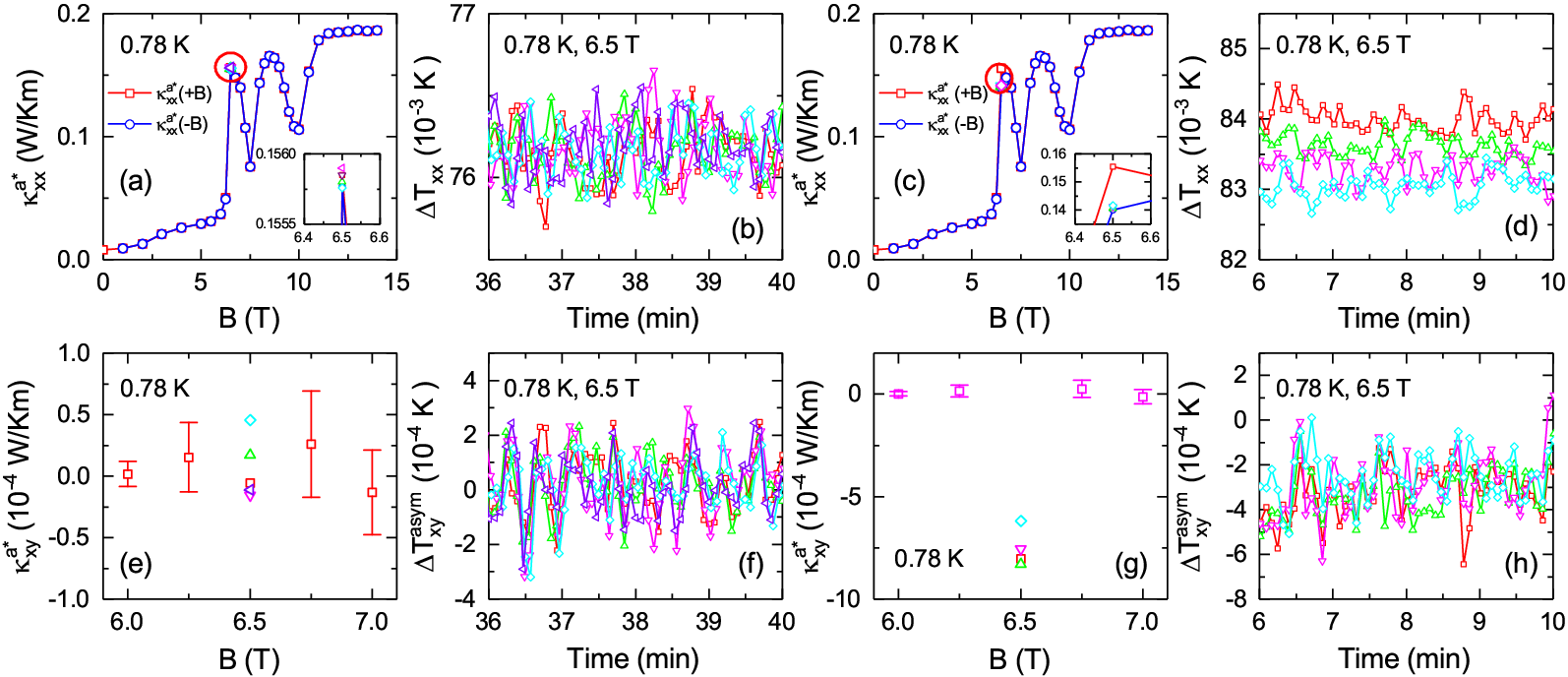}
\caption{(a,c) Magnetic field dependence of the longitudinal thermal conductivity $\kappa^{a*}_{xx}$ for $+B$ (red open squares) and $-B$ (other symbols) at 0.78 K. The insets in (a) and (c) show the data at $B_{a*1}$. (e,g) Thermal Hall conductivity $\kappa^{a*}_{xy}$ for $+B$ (red open squares) and $-B$ (other symbols) at 0.78 K. In panel (a), several repeated measurements gave indistinguishable $\kappa^{a*}_{xx}(B)$ values at $B_{a*1}$ and corresponding small $\kappa^{a*}_{xy}$ values, as indicated by the solid symbols in panel (e). However, in panel (c), several measurements gave small differences among $\kappa^{a*}_{xx}(B)$ at $B_{a*1}$ and corresponding large negative values, as indicated by the solid symbols in panel (g). (b,f) Longitudinal temperature difference $\Delta T_{xx}$ and transverse temperature difference $\Delta T^{asym}_{xy}$ as a function of time that correspond to panels (a) and (e), respectively. (d,h) Longitudinal temperature difference $\Delta T_{xx}$ and transverse temperature difference $\Delta T^{asym}_{xy}$ as a function of time that correspond to panels (c) and (g), respectively. The difference between the measurements of (a,e) and those of (c,g) is that for the former ones the data were recorded after long time ($>$ 30 minutes) waiting for achieving the steady state of heat flowing, while for the later ones the data were recorded after only several minutes. The resulted difference in the $\kappa^{a*}_{xy}$ and $\kappa^{a*}_{xy}$ is significant only at $B_{a*1}$.}
\end{figure*}

To clarify the most significant difference in the $\kappa^{a*}_{xy}(B)$ data at $B_{a*1}$ between two groups, we took very careful checks on the measurements and noticed that it may be related to the rather long relaxation time of temperature gradients at this critical field. For this checking measurement, particularly at $B_{a*1}$ and at 0.78 K, the temperature gradients were measured with different waiting times after applying the heat current. At this temperature and in 0--14 T fields, usually several minutes are enough to establish the equilibrium heat flowing state. However, at $B_{a*1}$ we found that the relaxation time is much longer. First, we carried out a set of measurement with long waiting time. Figure 3(a) shows the $\kappa^{a*}_{xx}(B)$ data for $+B$ and $-B$ fields and Fig. 3(e) shows the corresponding $\kappa^{a*}_{xx}(B)$ data, for which all the temperature gradients were recorded with long enough waiting time ($>$ 30 minutes) after applying the heat current. Figures 3(b) and 3(f) show the longitudinal and transverse temperature gradients, which are stable enough and only weakly fluctuating with flat based lines. In this case, the $\kappa^{a*}_{xx}$ data at -6.5 T (several times measurements) are indistinguishable from that at 6.5 T; accordingly, the obtained $\kappa^{a*}_{xy}$ data at 6.5 T are small and the $\kappa^{a*}_{xy}(B)$ curve does not show any anomaly at this critical field. Second, at -6.5 T the temperature gradients were recorded with usual waiting time (several minutes) after applying the heat current, as shown in Fig. 3(d) and 3(h). In this case, the temperature gradients are not stable enough although the shifts of temperature gradients are very weak. Accordingly, the $\kappa^{a*}_{xx}$ data at -6.5 T are slightly smaller than that at 6.5 T and the obtained $\kappa^{a*}_{xy}$ data show rather large negative values, as shown in Fig. 3(c) and 3(g). The later measurements actually are similar to the dip-like feature at $B_{a*1}$ observed by Takeda {\it et al.} \cite{Takeda}. Thus, it can be concluded that this dip-like feature is not intrinsic.

\begin{figure*}
\includegraphics[clip,width=18cm]{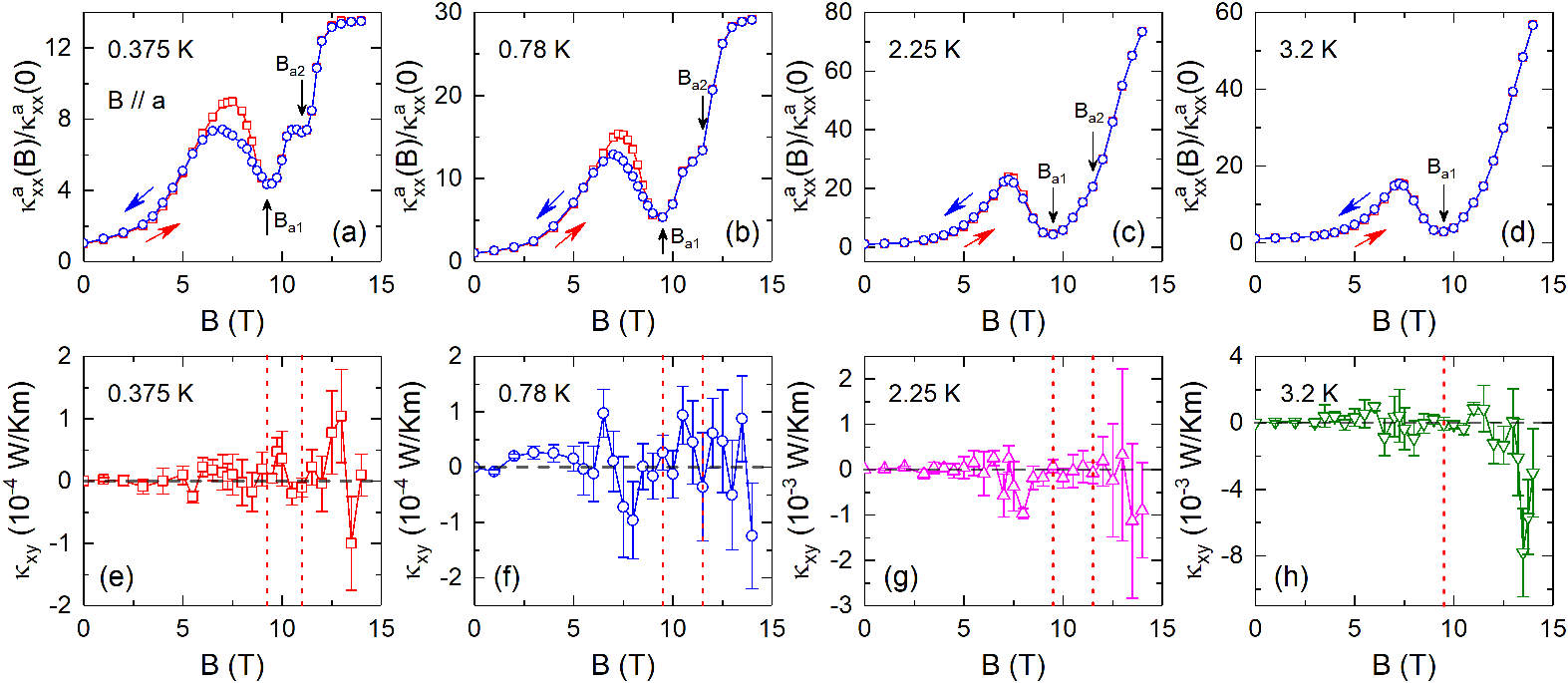}
\caption{Longitudinal thermal conductivity $\kappa_{xx}$ and thermal Hall conductivity $\kappa_{xy}$ of Na$_2$Co$_2$TeO$_6$ single crystals as a function of magnetic field with magnetic field and heat current applied along the $a$ axis. (a-d) Isotherms of the $\kappa_{xx}$ with respect to the zero-field value as a function of the magnetic field at 0.375, 0.78, 2.25 and 3.2 K, respectively. The red and blue arrows indicate the process of applying magnetic field, and an obvious hysteresis can be observed below $B_{a1}$. Black arrows indicate the critical fields as mentioned in main text. (e-h) Magnetic field dependence of thermal Hall conductivity $\kappa_{xy}$ at different selected temperatures at 0.375, 0.78, 2.25, and 3.2 K, respectively. At each temperature the measurements were repeated more than five times to get the averaged data and error bars. The red dot line indicates the critical fields of magnetic transitions.}
\end{figure*}

The planar thermal Hall conductivity and thermal conductivity were also measured with magnetic field and heat current along the $a$ axis. As shown in Figs. 4(a-d), the longitudinal thermal conductivity $\kappa^a_{xx}(B)$ exhibit the similar behaviors to the previous ultralow temperature data \cite{Guang}. With increasing field, the $\kappa^a_{xx}$ firstly increases and arrives a maximum at $\sim$ 7.5 T, and then decreases and reaches a minimum at $B_{a1} \sim$ 9.75 T, which is close to the spin polarization transition for $B \parallel a$. Above $B_{a1}$, the $\kappa^a_{xx}$ quickly increases and finally saturates in the polarized state. However, another kink-like anomaly appears in $\kappa^a_{xx}(B)$ at 10.25--10.5 T, which is marked as $B_{a2}$. This anomaly is likely due to a unexplored field-induced magnetic transition. Since this anomaly becomes indistinguishable at temperatures above 2.25 K, it cannot be probed by eralier magnetization studies \cite{Hong}. In addition, there is an obvious hysteresis between the increasing and decreasing field data for $B < B_{a1}$. The unusual hysteresis behavior for $B < B_{a1}$ may be related to the magnetic domains which can scatter phonons in Na$_2$Co$_2$TeO$_6$.

Figures 4(e-h) show the planar thermal Hall conductivity $\kappa^a_{xy}$ as a function of magnetic field at the selected temperatures. The data at $B < B_{a1}$ are rather similar at all temperatures, that is, there is weak positive signal at low fields and negative signal at the intermediate fields. A sign change of $\kappa^a_{xy}$ seems to occur at the peak field of $\kappa^a_{xx}(B)$. A peak-like behavior appears for $B_{a1} < B < B_{a2}$ at low temperatures and disappears at 2.25 and 3.2 K. Rather large signal was observed for $B > B_{a2}$: there is large positive signal accompanied with sign change at low temperatures, which evolutes into a large negative dip at 3.2 K. This high-field feature are rather similar between the present data and Takeda {\it et al.}'s data at low temperatures \cite{Takeda}. However, their data at high temperatures show positive signal for $B > B_{a2}$.

One of our main experimental results is that both the $a*$- and $a$-axis magnetic fields can induce planar magnetic Hall effect in Na$_2$Co$_2$TeO$_6$ at low temperatures. This essentially reproduces the previous results from Takeda {\it et al.} \cite{Takeda}, which were explained to be originated from the topological magnons. However, there are some obvious difference in the $\kappa_{xy}(B)$ between two groups' results. First, Takeda {\it et al.}'s results indicated that the planar Hall effect is closely correlated to the field-induced magnetic transitions. In contrast, our data indicate that the planar thermal Hall conductivity does not show up at the magnetic transition fields for both $B \parallel a*$ and $B \parallel a$. In particular, the most striking feature in Takeda {\it et al.}'s data is the sharp dip of $\kappa^{a*}_{xy}(B)$ data at $B_{a*1}$, which however was proved to be artificial by our careful checking on the relaxation time. This kind of difference suggests that the origin of the planar thermal Hall effect in Na$_2$Co$_2$TeO$_6$ needs further experimental investigations. Second, although both groups' data confirmed the quite large planar thermal Hall conductivity at high fields, there exist rather clear difference in the details like the sign change of $\kappa_{xy}(B)$. Another main result is that our thermal conductivity data $\kappa_{xx}(B)$ display much sharper anomalies at the field-induced magnetic transitions, which allows us to observe an additional kink-like anomaly in the $\kappa_{xx}(B)$ curves with $B \parallel a$. This indicates that there is likely another unexplored field-induced magnetic transition at $B_{a2}$. Therefore, the magnetic phase diagram of Na$_{2}$Co$_{2}$TeO$_{6}$ may be more complicated than what people have known.

Here we give more discussion about the indication of the experimental results of $\kappa_{xx}$ and $\kappa_{xy}$ for the fields along different in-plane directions. Unlike the $\kappa_{xx}$ result, the $\kappa_{xy}$ reveals the wavefunction properties or Berry curvature properties of the relevant heat carriers, such as the magnetic excitations and/or the phonon modes. While the $\kappa_{xy}$ is also sensitive to the density of carriers' states and their mutual scattering like $\kappa_{xx}$, the Berry curvature properties can bring extra features such as the large change in the magnitude and the sign reversal to the $\kappa_{xy}$. This phenomenon occurs when the bands of the excitations experience radical changes in their Berry curvature distributions. For example, the excited magnon bands touch and reopen with a significant change of their Chern numbers. Actually, this change does not have to be associated with a magnetic structure transition. The magnetic structure can be smoothly varied by the external magnetic field without experiencing any phase transition while the change of magnon band structure topology can occur with a large change in $\kappa_{xy}$ \cite{Neumann, Chen}. Thus, the $\kappa_{xy}$ provides more information about the properties of the excitations in the system. In Figs. 2 and 4, as we have previously mentioned, at the magnetic transitions that are indicated in the $\kappa_{xx}$, there are the associated changes in the $\kappa_{xy}$ if the signal is discernible. Nevertheless, the $\kappa_{xy}$ still experience sign reversals or large magnitude change at the magnetic fields where there are no obvious signatures in the $\kappa_{xx}$. This is clearly observed in Fig. 2 for $B > B_{a*3}$ and in Fig. 4 for $B > B_{a2}$. Thus, a radical change of the magnon band structure topology and the associated Berry curvature distribution may be experienced before the system enters a full polarized state. In fact, some recent theoretical researches have demonstrated that for the in-plane magnetic fields the thermal Hall conductivity of Kitaev magnets arises from the topological magnons with finite Chern numbers and a peculiar sign structure follows from the symmetries of the momentum space Berry curvature \cite{Chern, Zhang}. This scenario is compatible with our reasoning above, and may be adopted to describe the planar thermal Hall effect of Na$_2$Co$_2$TeO$_6$, particularly the pronounced $\kappa_{xy}(B)$ in high-field region.

In addition to the properties about the magnetic excitations and magnetic structures in the fields, our results are quite indicative about the candidate model for Na$_2$Co$_2$TeO$_6$. The $XXZ$ spin model that was recently proposed for the honeycomb cobalt BaCo$_2$(AsO$_4$)$_2$ has a global U(1) symmetry. This implies that the heat transport for the fields in the $a$ and $a*$ directions should behave identically. Our experiments for Na$_2$Co$_2$TeO$_6$, however, show quite distinct behaviors for these two directions. This strongly indicates that the global U(1) symmetry should be absent for Na$_2$Co$_2$TeO$_6$. We thus think more anisotropic spin interactions should be included for understanding the physics of Na$_2$Co$_2$TeO$_6$. These results put some constraints on the candidate models for Na$_2$Co$_2$TeO$_6$.

\section{Summary}

In summary, we study both the in-plane thermal conductivity and thermal Hall conductivity of Na$_2$Co$_2$TeO$_6$ in magnetic fields along the honeycomb layer. It is found that the zero-field thermal conductivity displays weak in-plane anisotropy, while the magnetic fields along the $a$ or $a*$ axis affect the $\kappa_{xx}$ quite differently. The field-induced magnetic transitions induce anomalies at the low-temperature $\kappa^a_{xx}(B)$ and $\kappa^{a*}_{xx}(B)$ curves. Although most of these transitions have been probed by various experiments including thermal conductivity, the present $\kappa^a_{xx}(B)$ data for $B \parallel a$ reveal an unexplored magnetic transition at very low temperatures. Either the $a$-axis or the $a*$-axis magnetic fields can induce planar thermal Hall effect, which is mainly related to the magnon Berry curvature rather than the field-induced magnetic transitions. These results put clear constraints on the in-plane high-field phase of Na$_2$Co$_2$TeO$_6$ and the anisotropic spin interactions must be included in theoretical models.

\begin{acknowledgments}

This work was supported by the National Natural Science Foundation of China (Grant Nos. 12274388, 12174361, 12104010, and 12104011), the Nature Science Foundation of Anhui Province (Grant Nos. 1908085MA09 and 2108085QA22), and the Research Grants Council of Hong Kong with C7012-21GF. The work at the University of Tennessee was supported by the NSF with Grant No. NSF-DMR-2003117.

\end{acknowledgments}

\end{document}